\documentclass[preprint,12pt]{elsarticle}  
\usepackage{graphicx} 
\usepackage{amssymb} 
\usepackage{amsmath} 

\journal{Physics Letters B} 

\begin{document} 

\begin{frontmatter} 

\title{Shell-model predictions for $\Lambda\Lambda$ 
hypernuclei} 

\author[a]{A.~Gal\corref{cor1}} 
\cortext[cor1]{Corresponding author: Avraham Gal, avragal@vms.huji.ac.il} 
\author[b]{D.J.~Millener} 
\address[a]{Racah Institute of Physics, The Hebrew University, 91904 
Jerusalem, Israel} 
\address[b]{Physics Department, Brookhaven National Laboratory, Upton, 
NY 11973, USA} 

\date{\today} 

\begin{abstract} 
It is shown how the recent shell-model determination of $\Lambda N$ spin 
dependent interaction terms in $\Lambda$ hypernuclei allows for a reliable 
deduction of $\Lambda\Lambda$ separation energies in $\Lambda\Lambda$ 
hypernuclei across the nuclear $p$ shell. Comparison is made with the 
available data, highlighting $_{\Lambda\Lambda}^{~11}{\rm Be}$ and 
$_{\Lambda\Lambda}^{~12}{\rm Be}$ which have been suggested as possible 
candidates for the KEK-E373 Hida event. 
\end{abstract} 


\end{frontmatter} 

\section{Introduction} 
\label{sec:intro} 

The properties of hypernuclei reflect the nature of the underlying 
baryon-baryon interactions and, thus, provide useful information on the 
in-medium hyperon-nucleon and hyperon-hyperon interactions. Knowledge of 
these interactions is required to extrapolate into strange hadronic matter 
\cite{sbg93} for both finite systems and in bulk, and into neutron stars 
\cite{jsb08}. Whereas a fair amount of data is available on single-$\Lambda$ 
hypernuclei, including production, structure and decay modes \cite{hashim06}, 
little is known definitively on strangeness ${\cal S}=-2$ hypernuclei produced 
to date by recording $\Xi^-$ capture events in emulsion and following their 
decay sequence \cite{nakazawa10}. Normally these observed events do not offer 
unique assignments, except for $_{\Lambda\Lambda}^{~~6}{\rm He}$ which is the 
lightest particle-stable $\Lambda\Lambda$ hypernucleus established firmly 
so far \cite{takahashi01}. Numerous $_{\Lambda\Lambda}^{~~6}{\rm He}$ 
calculations have been reported since then, 
including Faddeev \cite{fg02a,fg02b} and variational \cite{hiyama02} 
$\alpha\Lambda\Lambda$ cluster calculations, as well as 
VMC \cite{shoeb04,ubs04} and stochastic variational \cite{nemura05,uh06} 
six-body calculations. The separation energy $B_{\Lambda\Lambda}$ 
of the two $\Lambda$'s in this system exceeds the sum of separation 
energies $B_{\Lambda}$ of each of its $\Lambda$'s in the single-$\Lambda$ 
hypernucleus $_{\Lambda}^{5}{\rm He}$ by less than 1 MeV \cite{nakazawa10}: 
\begin{equation} 
\Delta B_{\Lambda\Lambda}(_{\Lambda\Lambda}^{~~6}{\rm He}) \equiv 
B_{\Lambda\Lambda}(_{\Lambda\Lambda}^{~~6}{\rm He}) - 
2B_{\Lambda}(_{\Lambda}^{5}{\rm He}) = 0.67\pm 0.17~{\rm MeV}. 
\label{eq:LL6He} 
\end{equation} 
Owing to the weakness of the $\Lambda\Lambda$ interaction, it is reasonable 
to identify $\Delta B_{\Lambda\Lambda}(_{\Lambda\Lambda}^{~~6}{\rm He})$ 
with the $^1S_0$ $\Lambda\Lambda$ interaction shell-model (SM) matrix element 
$<V_{\Lambda\Lambda}>_{\rm SM}$ in the $(1s_{\Lambda})^2$ ground state (g.s.) 
configuration of neighboring hypernuclei. This argument suggests the following 
estimate for $B_{\Lambda\Lambda}$ in the nuclear $p$ shell: 
\begin{equation} 
B_{\Lambda\Lambda}^{\rm SM}(_{\Lambda\Lambda}^{~~\rm A}{\rm Z})=
2{\overline B}_{\Lambda}({_{~~~\Lambda}^{\rm A-1}{\rm Z}}) + 
<V_{\Lambda\Lambda}>_{\rm SM}, 
\label{eq:BLL} 
\end{equation} 
where ${\overline B}_{\Lambda}({_{~~~\Lambda}^{\rm A-1}{\rm Z}})$ is 
the $(2J+1)$-averaged $B_{\Lambda}$ in the single-$\Lambda$ hypernucleus 
$_{~~~\Lambda}^{\rm A-1}{\rm Z}$ g.s. doublet, as appropriate to a spin 
zero $(1s_{\Lambda})^2$ configuration of the double-$\Lambda$ hypernucleus 
$_{\Lambda\Lambda}^{~~\rm A}{\rm Z}$ \cite{fg02b,hiyama02}, and 
$<V_{\Lambda\Lambda}>_{\rm SM}=0.67\pm 0.17~{\rm MeV}$.{\footnote{In cluster 
model (CM) calculations \cite{hiyama10}, $<V_{\Lambda\Lambda}>_{\rm CM} \equiv 
B_{\Lambda\Lambda}(V_{\Lambda\Lambda}\neq 0)-B_{\Lambda\Lambda}
(V_{\Lambda\Lambda}=0)$ assumes similar values: 0.54 MeV for 
$_{\Lambda\Lambda}^{~~6}{\rm He}$ and 0.53 MeV for 
$_{\Lambda\Lambda}^{~10}{\rm Be}$.}} 

The remarkably simple SM estimate (\ref{eq:BLL}), for nuclear core spin 
$J_c \neq 0$, requires the knowledge of $B_{\Lambda}$ for both the g.s. as 
well as its $\Lambda$ hypernuclear doublet partner which normally is the 
first excited state and which experimentally is not always known. However, 
recent advances in $\Lambda$ hypernuclear spectroscopy \cite{tamura10} made 
it possible to derive the spin-dependent $\Lambda N$ interaction matrix 
elements \cite{mil10} directly from $\gamma$ ray measurements, and thus 
to relate ${\overline B}_{\Lambda}({_{~~~\Lambda}^{\rm A-1}{\rm Z}})$ 
to experimentally determined g.s. separation energies 
$B_{\Lambda}^{\rm exp}({_{~~~\Lambda}^{\rm A-1}{\rm Z}_{\rm g.s.}})$. 
In this Letter, we discuss briefly the g.s. spectroscopy of single-$\Lambda$ 
hypernuclei and show how to derive the ${\overline B}_{\Lambda}$ input to 
Eq.~(\ref{eq:BLL}) in order to predict g.s. $B_{\Lambda\Lambda}$ values in 
$\Lambda\Lambda$ hypernuclei. These predictions work remarkably well for 
$_{\Lambda\Lambda}^{~10}{\rm Be}$ and $_{\Lambda\Lambda}^{~13}{\rm B}$ which 
are the only emulsion events assigned with high degree of consistency so 
far beyond $_{\Lambda\Lambda}^{~~6}{\rm He}$. We present predictions for 
$_{\Lambda\Lambda}^{~11}{\rm Be}$, $_{\Lambda\Lambda}^{~12}{\rm Be}$ and 
$_{\Lambda\Lambda}^{~12}{\rm B}$ to confront several recently reported 
alternative interpretations to the KEK-E176 event generally accepted as 
$_{\Lambda\Lambda}^{~13}{\rm B}$ \cite{aoki09}, and particularly to confront 
the very recent KEK-E373 HIDA event \cite{nakazawa10}. Our SM prediction 
for $_{\Lambda\Lambda}^{~11}{\rm Be}$ agrees with a recent CM prediction 
for $_{\Lambda\Lambda}^{~11}{\rm Be}$ treated as a five-body 
$\alpha\alpha n \Lambda\Lambda$ cluster \cite{hiyama10}. Our SM prediction 
for $_{\Lambda\Lambda}^{~12}{\rm Be}$ is without competition, resulting in 
a definitive statement that neither $_{\Lambda\Lambda}^{~11}{\rm Be}$ nor 
$_{\Lambda\Lambda}^{~12}{\rm Be}$ provides satisfactory agreement with the 
HIDA emulsion event \cite{nakazawa10}.

\section{Input from $\Lambda$ hypernuclear shell model} 
\label{sec:SM} 

The $\Lambda N$ effective interaction for a $1s_{\Lambda}$ orbital in the 
nuclear $p$ shell is given by \cite{gsd71} 
\begin{equation} 
V_{\Lambda N}(r)  =  V_0(r) + V_{\sigma}(r)~ \vec{s}_N\cdot 
\vec{s}_{\Lambda} +  V_{\Lambda }(r)~\vec{l}_{N\Lambda }\cdot 
\vec{s}_{\Lambda}  + V_{\rm N}(r)~\vec{l}_{N \Lambda }\cdot 
\vec{s}_{N} +  V_{\rm T}(r)~S_{12}\; , 
\label{eq:vlam} 
\end{equation} 
where $S_{12} = 3(\vec{\sigma}_{N}\cdot\vec{r})(\vec{\sigma}_{\Lambda} 
\cdot\vec{r})-\vec{\sigma}_{N}\cdot\vec{\sigma}_{\Lambda}$. The five 
$p_N s_\Lambda$ two-body matrix elements depend on the radial integrals 
associated with each component in Eq.~(\ref{eq:vlam}). They are denoted 
by the parameters $\overline{V}$, $\Delta$, $S_\Lambda$, $S_N$ and $T$. 
By convention, $S_\Lambda$ and $S_N$ are actually the coefficients 
of $\vec{l}_N\cdot\vec{s}_\Lambda$ and $\vec{l}_N\cdot\vec{s}_N$. 
Then, the operators associated with $\Delta$ and $S_\Lambda$ are 
$\vec{S}_N\cdot \vec{s}_{\Lambda}$ and $\vec{L}_{N}\cdot \vec{s}_{\Lambda}$. 
We note that $\overline{V}$ contributes only to the overall binding energy 
and $S_N$ does not contribute to doublet splittings in the weak-coupling 
limit, but augments the nuclear spin-orbit interaction and contributes 
to the spacings between states based on different core states. The 
doublet splittings are determined to an excellent approximation by 
the $\Lambda$ spin-dependence parameters $\Delta$, $S_\Lambda$ and $T$, 
together with important contributions from $\Lambda\!-\!\Sigma$ coupling. 

The parametrization of Eq.~(\ref{eq:vlam}) applies to the direct $\Lambda N$ 
interaction, the $\Lambda N$--$\Sigma N$ coupling interaction, and the direct 
$\Sigma N$ interaction for both isospin 1/2 and 3/2. A set of parameters that 
fits the full particle-stable excitation spectra of $_{\Lambda}^{7}{\rm Li}$ 
and $_{\Lambda}^{9}{\rm Be}$, six levels in total determined by the seven 
observed $\gamma$ rays \cite{tamura10}, is given by 
\begin{equation} 
\Delta =0.430\quad S_\Lambda =-0.015\quad {S}_{N} =-0.390\quad {T}=0.030\quad 
(A\!=\!7\!-\!9) \; , 
\label{eq:param7} 
\end{equation} 
all in MeV \cite{mil10}. Somewhat reduced values were derived in the heavier 
$p$-shell hypernuclei by fitting $\Delta$, $S_\Lambda$ and $T$ to the six 
$\Lambda$ hypernuclear doublet splittings deduced from $\gamma$ rays observed 
in mass $A\!=\!11\!-\!16$ hypernuclei \cite{tamura10}, and by fixing the 
parameter $S_N$ from the excitation energy of 
$_{~\Lambda}^{16}{\rm O}(1_{2}^{-})$: 
\begin{equation} 
\Delta =0.330\quad S_\Lambda =-0.015\quad {S}_{N} =-0.350\quad {T}=0.024\quad 
(A\!=\!11\!-\!16) \; , 
\label{eq:param11} 
\end{equation} 
all in MeV \cite{mil10}.
The corresponding matrix elements for $\Lambda\!-\!\Sigma$ coupling, based 
on $G$-matrix calculations \cite{akaishi00} for the NSC97e,f interactions 
\cite{rijken99}, are kept fixed throughout the $p$ shell \cite{mil10}: 
\begin{equation} 
\overline{V}' = 1.45\quad \Delta'= 3.04\quad S_\Lambda' = S_N' = -0.09
\quad T' = 0.16\quad  \,\,\, ({\rm in~MeV}). 
\label{eq:paramls} 
\end{equation} 

\begin{table}[hbt] 
\begin{center} 
\caption{$\Lambda\!-\!\Sigma$ and spin-dependent contributions to g.s. 
separation energies $B_{\Lambda}$ used for input to the calculation of 
$\Lambda\Lambda$ separation energies $B_{\Lambda\Lambda}$. 
The spin-independent term $\overline{V}$ is obtained (see text) from 
$B_{\Lambda}^{\rm exp}$ values \cite{davis05}. The $B_{\Lambda}^{\rm exp}$ 
values listed for the $A=9,10$ charge-symmetric doublets are statistical 
averages of the separate values, whereas the listed value for $A=12$ is 
$B_{\Lambda}^{\rm exp}({^{12}_{~\Lambda}{\rm B}})$.} 
\begin{tabular}{ccccccc} 
\hline 
${_{\Lambda}^{\rm A}{\rm Z}}$ & $^{9}_\Lambda$Be & 
$^{9}_\Lambda$Li--$^{9}_\Lambda$B & $^{10}_{~\Lambda}$Be--$^{10}_{~\Lambda}$B 
& $^{11}_{~\Lambda}$B & $^{11}_{~\Lambda}$Be & 
$^{12}_{~\Lambda}$B--$^{12}_{~\Lambda}$C \\ 
$J^{\pi}(\rm g.s.)$ & $1/2^+$ & $3/2^+$ & $1^-$ & $5/2^+$ & $1/2^+$ & $1^-$ \\ 
\hline 
$\Lambda\!-\!\Sigma$ & 4 & 183 & 35 & 66 & 99 & 103 \\ 
$\Delta$ & 0 & 350 & 125 & 203 & 2 & 108 \\ 
$S_\Lambda$ & 0 & $-10$ & $-13$ & $-20$ & 0 & $-14$ \\
$S_N$ & 207 & 434 & 386 & 652 & 540 & 704 \\ 
$T$ & 0 & $-6$ & $-15$ & $-43$ & 0 & $-29$ \\ 
\hline 
Sum (keV) & 211 & 952 & 518 & 858 & 641 & 869 \\ 
\hline 
$B_{\Lambda}^{\rm exp}$ (MeV) & 6.71 & 8.44 & 8.94 & 10.24 & & 11.37 \\ 
Error & $\pm$0.04 & $\pm$0.10 & $\pm$0.11 & $\pm$0.05 &  & $\pm$0.06 \\ 
\hline 
$\overline{V}$ (MeV) & $-0.84$ & $-1.09$ & $-1.06$ & $-1.04$ & & $-1.05$ \\ 
\hline 
\end{tabular} 
\label{tab:L} 
\end{center} 
\end{table} 

In Table \ref{tab:L} we list the $\Lambda\!-\!\Sigma$ and spin-dependent 
contributions to g.s. separation energies $B_{\Lambda}$ of interest in 
the present work, calculated in the shell model \cite{mil10}. Subtracting 
these contributions plus an $s_\Lambda$ single-particle energy identified 
with $B_{\Lambda}^{\rm exp}(^{5}_\Lambda{\rm He})\!=\!3.12\pm 0.02$ MeV, 
and dividing by the number of $p$-nucleons $(A\!-\!5)$, we obtain the 
magnitude of the spin-independent $s_{\Lambda}p_N$ matrix element 
$\overline{V}$. Our error estimate for $\overline{V}$ is $\pm 0.03$ MeV. 
Excluding $^{9}_\Lambda{\rm Be}$ which deviates substantially from the 
other species, a common value of $\overline{V}^{\rm SM}\!=\!-1.06\pm 0.03$ 
MeV emerges. The constancy of $\overline{V}$ in this mass range provides, 
a-posteriori, justification of the SM approach adopted here for $\Lambda$ 
hypernuclear g.s. studies. We can immediately check whether $\overline{V}$ 
is globally compatible also with the SM description of $B_{\Lambda\Lambda}$ 
within the range of experimentally known $\Lambda\Lambda$ hypernuclei, from 
$_{\Lambda\Lambda}^{~~6}{\rm He}$ to $_{\Lambda\Lambda}^{~13}{\rm B}$. 
To this end we form the difference between the $B_{\Lambda\Lambda}^{\rm exp}$ 
values, subtract 1.54 MeV for the contribution of $\Lambda \! - \! \Sigma$ 
coupling and $S_N$ to $B_{\Lambda\Lambda}({_{\Lambda\Lambda}^{~13}{\rm B}})$ 
(the contributions from the $\Lambda$-spin-dependent parameters 
$\Delta$, $S_\Lambda$ and $T$ average to zero in ${\overline B}_{\Lambda}$) 
and divide by $2\times (13-6)=14$ for the coefficient of $\overline{V}$ in 
this difference, see Eq.~(\ref{eq:BLL}). This gives 
$\overline{V}^{\rm SM}\!=\!-1.06\pm 0.05$ MeV, in excellent agreement 
with the value derived above from $\Lambda$ hypernuclear systematics. 

The departure of $^{9}_\Lambda{\rm Be}$ from the $\overline{V}$ systematics 
in Table \ref{tab:L} deserves discussion because 
$B_{\Lambda}(^{9}_\Lambda{\rm Be}_{\rm g.s.})$ has been a problem for SM 
studies of hypernuclei \cite{gsd71,mgdd85}. It is a somewhat artificial 
problem because, by convention, $B_{\Lambda}(^{9}_\Lambda{\rm Be})$ and 
$B_{\Lambda\Lambda}(_{\Lambda\Lambda}^{~10}{\rm Be})$ 
involve the binding energy of the free $^8$Be. Two $\alpha$'s just bound 
in a $2s$ relative state have a large separation. However, it takes only 
a few MeV of binding energy, such as the 3.5 MeV $\alpha$ separation energy 
in $^{9}_\Lambda{\rm Be}$, for the system to be reduced to a typical $p$-shell 
size. The $^5_\Lambda{\rm He}\!+\!\alpha$ system in a $1d$ relative state has 
a comparable radius even for a small $\alpha$ binding energy because of the 
additional centrifugal barrier. Therefore, we argue that the $\Lambda$'s in 
$^{9}_\Lambda{\rm Be}$ and $_{\Lambda\Lambda}^{~10}{\rm Be}$ interact within 
a nuclear core which is already of a normal $p$-shell size. By taking one 
$\overline{V}$ from $^{9}_\Lambda{\rm Be}$ and one from the 
$^{9}_\Lambda$Li--$^{9}_\Lambda$B column in Table~\ref{tab:L}, we incorporate 
the influence of the free $^8$Be binding energy into our estimate for 
$B_{\Lambda\Lambda}({_{\Lambda\Lambda}^{~10}{\rm Be}})$. In a slight 
variation, we can replace the contribution of the spin-independent component 
of the $\Lambda N$ interaction to one of the ${\overline B}_{\Lambda}$ values 
in Eq.~(\ref{eq:BLL}) by its average value $\overline{V}^{\rm SM}$, leading to 
\begin{equation} 
B_{\Lambda\Lambda}^{\rm SM}(_{\Lambda\Lambda}^{~~\rm A}{\rm Z})= 
2{\overline B}_{\Lambda}({_{~~~\Lambda}^{\rm A-1}{\rm Z}}) + 
(A-6)[\overline{V}({_{~~~\Lambda}^{\rm A-1}{\rm Z}})-\overline{V}^{\rm SM}] + 
<V_{\Lambda\Lambda}>_{\rm SM}, 
\label{eq:BLLalt} 
\end{equation} 
where $\Lambda\!-\!\Sigma$ contributions $\lesssim 0.1$ MeV are disregarded. 
The cluster model, on the other hand, has the freedom to treat $^8$Be itself 
as well as the hypernuclear systems. 

\section{Shell-model calculations and predictions for $\Lambda\Lambda$ 
hypernuclei} 
\label{sec:LL} 

In this section we discuss {\it all} the $\Lambda\Lambda$ hypernuclear 
species known or conjectured beyond $_{\Lambda\Lambda}^{~~6}{\rm He}$, 
the latter serving in both the SM and CM to constrain the $\Lambda\Lambda$ 
interaction. These consist of the accepted $_{\Lambda\Lambda}^{~10}{\rm Be}$, 
$_{\Lambda\Lambda}^{~13}{\rm B}$, two KEK-E176 event interpretations that 
could replace the generally accepted $_{\Lambda\Lambda}^{~13}{\rm B}$ 
interpretation \cite{aoki09}, and the recently published KEK-E373 HIDA 
emulsion event \cite{nakazawa10} assigned as $_{\Lambda\Lambda}^{~11}{\rm Be}$ 
or $_{\Lambda\Lambda}^{~12}{\rm Be}$. All are listed in Table~\ref{tab:LL}. 
For $_{\Lambda\Lambda}^{~10}{\rm Be}_{\rm g.s.}$, we follow Hiyama et 
al. \cite{hiyama02} assuming that the Demachi-Yanagi event \cite{ahn01} 
corresponds to the formation of $_{\Lambda\Lambda}^{~10}{\rm Be}^*_{2^+}
(3~{\rm MeV})$. The older Danysz event \cite{danysz63}, when fitted to 
a $\pi^-$ decay of $_{\Lambda\Lambda}^{~10}{\rm Be}_{\rm g.s.}$ to 
$_{\Lambda}^{9}{\rm Be}^*(3~{\rm MeV})$, yields $B_{\Lambda\Lambda}\!=\!
14.7\pm 0.4$ MeV, consistent with the value listed in the table. 
For $_{\Lambda\Lambda}^{~13}{\rm B}$, observed in KEK-E176 \cite{aoki91}, 
we follow the recent E4 event classification in Ref.~\cite{aoki09} assuming 
a $\pi^-$ decay of $_{\Lambda\Lambda}^{~13}{\rm B}_{\rm g.s.}$ to 
$_{~\Lambda}^{13}{\rm C}^*(4.9~{\rm MeV})$. These particular identifications 
for $_{\Lambda\Lambda}^{~10}{\rm Be}$ and $_{\Lambda\Lambda}^{~13}{\rm B}$, 
ensure that $0 \leq \Delta B_{\Lambda\Lambda}\lesssim 1$~MeV in both, 
similarly to Eq.~(\ref{eq:LL6He}) for $_{\Lambda\Lambda}^{~~6}{\rm He}$. 

\begin{table}[hbt] 
\begin{center} 
\caption{Comparison between $B_{\Lambda\Lambda}^{\rm exp}$ from KEK-E176,E373 
and $B_{\Lambda\Lambda}^{\rm SM}({_{\Lambda\Lambda}^{~~\rm A}{\rm Z}})$ 
predictions using Eq.~(\ref{eq:BLL}) [Eq.~(\ref{eq:BLLalt}) for 
$_{\Lambda\Lambda}^{~10}{\rm Be}$]. Input ${\overline B}_{\Lambda}$ values, as 
well as $B_{\Lambda\Lambda}^{\rm CM}({_{\Lambda\Lambda}^{~~\rm A}{\rm Z}})$
values \cite{hiyama10} where available, are also listed. All values are in 
MeV. The $B_{\Lambda\Lambda}^{\rm exp}$ values from KEK-E176 \cite{aoki09} 
refer to different interpretations of the same emulsion event.}
\begin{tabular}{ccccc} 
\hline 
~$_{\Lambda\Lambda}^{~~\rm A}{\rm Z}$~ & 
~${\overline B}_{\Lambda}({_{~~~\Lambda}^{\rm A-1}{\rm Z}})$~ & 
~$B_{\Lambda\Lambda}^{\rm SM}({_{\Lambda\Lambda}^{~~\rm A}{\rm Z}})$~ & 
~$B_{\Lambda\Lambda}^{\rm exp}({_{\Lambda\Lambda}^{~~\rm A}{\rm Z}})$~ &
~$B_{\Lambda\Lambda}^{\rm CM}({_{\Lambda\Lambda}^{~~\rm A}{\rm Z}})$~ \\
\hline 
~$_{\Lambda\Lambda}^{~~6}{\rm He}$~ & ~$3.12\pm 0.02$~ & ~$6.91\pm 0.16$~ & 
~$6.91\pm 0.16$~\cite{nakazawa10} & ~$6.91$~ \\ 
~$_{\Lambda\Lambda}^{~10}{\rm Be}$~ & ~$6.71\pm 0.04$~ & ~$14.97\pm 0.22$~ & 
~$14.94\pm 0.13$~ \cite{ahn01} & ~$14.74$~ \\
~$_{\Lambda\Lambda}^{~11}{\rm Be}$~ & ~$8.86\pm 0.11$~ & ~$18.40\pm 0.28$~ & 
~$17.53\pm 0.71$~\cite{aoki09} & ~$18.23$~ \\ 
 &  &  & ~$20.83\pm 1.27$~\cite{nakazawa10} &  \\ 
~$_{\Lambda\Lambda}^{~12}{\rm Be}$~ & ~$10.02\pm 0.05$~ & ~$20.72\pm 0.20$~ & 
~$22.48\pm 1.21$~\cite{nakazawa10} & --~  \\ 
~$_{\Lambda\Lambda}^{~12}{\rm B}$~ & ~$10.09\pm 0.05$~ & ~$20.85\pm 0.20$~ &
~$20.60\pm 0.74$~\cite{aoki09} & --~  \\ 
~$_{\Lambda\Lambda}^{~13}{\rm B}$~ & ~$11.27\pm 0.06$~ & ~$23.21\pm 0.21$~ & 
~$23.3\pm 0.7$~\cite{aoki09} & --~  \\ 
\hline 
\end{tabular} 
\label{tab:LL} 
\end{center} 
\end{table} 

$B_{\Lambda\Lambda}^{\rm SM}({_{\Lambda\Lambda}^{~\rm A}{\rm Z}})$ predictions 
obtained by applying Eq.~(\ref{eq:BLL}) [Eq.~(\ref{eq:BLLalt}) for 
$_{\Lambda\Lambda}^{~10}{\rm Be}$] are listed in Table~\ref{tab:LL} together 
with the ${\overline B}_{\Lambda}({_{~~~\Lambda}^{
\rm A-1}{\rm Z}})$ input which is constrained by known $B_{\Lambda}^{\rm exp}(
{_{~~~\Lambda}^{\rm A-1}{\rm Z}})$ values \cite{davis05}. A brief discussion 
is due. 
\begin{itemize} 
\item 
In the $B_{\Lambda\Lambda}({_{\Lambda\Lambda}^{~10}{\rm Be}})$ calculation, 
to account for the loose structure of the $^8$Be core, we used 
Eq.~(\ref{eq:BLLalt}) with $\overline{V}^{\rm SM}\!=\!-1.06\pm 0.03$ MeV. 
The calculated $B_{\Lambda\Lambda}^{\rm SM}$ provides excellent agreement 
within errors with the experimental value, and departs within 
$\approx 1\sigma$ from the CM calculated value. While Eq.~(\ref{eq:BLLalt}) 
uses the $p$-shell {\it average} $\Lambda N$ matrix element 
$\overline{V}^{\rm SM}$, it is possible alternatively to proceed locally 
focusing on the $A\!=\!9$ $\Lambda$ hypernuclei. In this procedure we replace 
one of the two ${\overline B}_{\Lambda}({_{\Lambda}^{9}{\rm Be}})$ in 
Eq.~(\ref{eq:BLL}) by an appropriate contribution from the more compact 
$^{9}_{\Lambda}{\rm Li}$ and $^{9}_{\Lambda}{\rm B}$ hypernuclei. 
Specifically, we subtract the `sum' entry 
of 952 keV in Table~\ref{tab:L} from $B_{\Lambda}^{\rm exp}(^{9}_{\Lambda}{\rm 
Li}-{^{9}_{\Lambda}{\rm B}})\!=\!8.44$ MeV there, adding the proper `sum' 
entry of 211 keV for $_{\Lambda}^{9}{\rm Be}$, which results in 7.70 MeV. The 
appropriate ${\overline B}_{\Lambda}$ is then given by the average of 6.71 MeV 
for the first $\Lambda$, as in $_{\Lambda}^{9}{\rm Be}$, and 7.70 MeV for the 
second $\Lambda$. This procedure results in $B_{\Lambda\Lambda}^{\rm SM}({_{
\Lambda\Lambda}^{~10}{\rm Be}})\!=\!15.08\pm 0.20$ MeV, in agreement with 
$B_{\Lambda\Lambda}^{\rm exp}({_{\Lambda\Lambda}^{~10}{\rm Be}})$ within error 
bars. 
\item 
In the $B_{\Lambda\Lambda}({_{\Lambda\Lambda}^{~11}{\rm Be}})$ calculation, 
${\overline B}_{\Lambda}({_{~\Lambda}^{10}{\rm Be}})$ was derived from a 120 
keV g.s. doublet splitting provided by the $\Lambda N$ interaction parameters 
(\ref{eq:param11}) and (\ref{eq:paramls}). The corresponding $2^- \to 1^-$ 
$\gamma$ ray transition has not been observed. One reason could be that the 
splitting is considerably smaller, causing the excited state to undergo weak 
decay. In the extreme case of degenerate doublet levels, using 
${\overline B}_{\Lambda}\!=\!B_{\Lambda}^{\rm exp}
({_{~\Lambda}^{10}{\rm Z}_{\rm g.s.}})\!=\!8.94$ MeV, the 
calculated $B_{\Lambda\Lambda}^{\rm SM}({_{\Lambda\Lambda}^{~11}{\rm Be}})$ 
would increase by 0.15 MeV with respect to the value listed in the table. 
We note that the SM prediction and the CM prediction agree with each other 
within error bars in spite of different $\Lambda N$ interaction inputs. This 
agreement might be fortuitous. Of the two experimental $B_{\Lambda\Lambda}$ 
assignments, only the KEK-E176 \cite{aoki09} G2 alternative assignment to 
$_{\Lambda\Lambda}^{~13}{\rm B}$ is in rough agreement within errors with the 
theoretical predictions. 
\item 
In the $B_{\Lambda\Lambda}({_{\Lambda\Lambda}^{~12}{\rm Be}})$ calculation, 
we subtracted the specific `sum' of 858 keV in Table~\ref{tab:L} from 
$B_{\Lambda}^{\rm exp}({_{~\Lambda}^{11}{\rm B}})\!=\!10.24$ MeV there, adding 
the proper `sum' of 641 keV for $_{~\Lambda}^{11}{\rm Be}_{\rm g.s.}$ to 
obtain ${\overline B}_{\Lambda}({_{~\Lambda}^{11}{\rm Be}})$. The calculated 
$B_{\Lambda\Lambda}^{\rm SM}({_{\Lambda\Lambda}^{~12}{\rm Be}})$ disagrees 
with a KEK-E373 Hida event assignment. 
\item 
In the $B_{\Lambda\Lambda}({_{\Lambda\Lambda}^{~12}{\rm B}})$ calculation, 
${\overline B}_{\Lambda}({_{~\Lambda}^{11}{\rm B}})$ was derived from 
the observed 263 keV g.s. doublet splitting \cite{ma10}. The predicted 
$B_{\Lambda\Lambda}$ value is in good agreement with the 
KEK-E176 \cite{aoki09} G3 alternative assignment to 
$B_{\Lambda\Lambda}({_{\Lambda\Lambda}^{~13}{\rm B}})$. 
\item 
In the $B_{\Lambda\Lambda}({_{\Lambda\Lambda}^{~13}{\rm B}})$ calculation, 
since the value of 
$B_{\Lambda}^{\rm exp}({_{~\Lambda}^{12}{\rm C}_{\rm g.s.}})$ is based on 
only 6 events and is unsettled, we used the much better determined 
$B_{\Lambda}^{\rm exp}({_{~\Lambda}^{12}{\rm B}_{\rm g.s.}})$ \cite{davis05} 
plus a 161 keV g.s. doublet splitting from the observed $2^-\to 1^-$ $\gamma$ 
ray transition in the charge-symmetric hypernucleus $_{~\Lambda}^{12}{\rm C}$ 
\cite{ma10}. 
\end{itemize} 

The very good agreement in Table~\ref{tab:LL} between 
$B_{\Lambda\Lambda}^{\rm SM}({_{\Lambda\Lambda}^{~10}{\rm Be}})$ and 
$B_{\Lambda\Lambda}^{\rm exp}({_{\Lambda\Lambda}^{~10}{\rm Be}})$, and 
between $B_{\Lambda\Lambda}^{\rm SM}({_{\Lambda\Lambda}^{~13}{\rm B}})$ 
and $B_{\Lambda\Lambda}^{\rm exp}({_{\Lambda\Lambda}^{~13}{\rm B}})$, 
provides a consistency check from below and from above on the predicted 
values in between. Our calculations suggest that the KEK-E176 \cite{aoki09} 
event interpretations G2 for $_{\Lambda\Lambda}^{~11}{\rm Be}$ and G3 for 
$_{\Lambda\Lambda}^{~12}{\rm B}$, with $B_{\Lambda\Lambda}$ values listed 
in Table~\ref{tab:LL}, cannot be excluded.{\footnote{An alternative 
production reaction to G3, specified by E2 and resulting in 
$B_{\Lambda\Lambda}({_{\Lambda\Lambda}^{~12}{\rm B}})=20.02\pm 0.78$ MeV 
\cite{aoki09}, also admits a $_{\Lambda\Lambda}^{~12}{\rm B}$ assignment.}} 
In contrast, it is difficult to reconcile the KEK-E373 Hida event 
interpretation as either $_{\Lambda\Lambda}^{~11}{\rm Be}$ or 
$_{\Lambda\Lambda}^{~12}{\rm Be}$ \cite{nakazawa10} with the calculated 
$B_{\Lambda\Lambda}^{\rm SM}$ listed in the table. We conclude that while 
a $_{\Lambda\Lambda}^{~12}{\rm Be}$ assignment of the HIDA event is somewhat 
more likely than a $_{\Lambda\Lambda}^{~11}{\rm Be}$ assignment, both 
assignments are ruled out by the SM.

\section{Discussion and Conclusion} 
\label{concl} 

Historically, cluster models charted the $B_{\Lambda\Lambda}$ 
map of $\Lambda\Lambda$ hypernuclei in the $p$ 
shell \cite{fg02a,fg02b,hiyama02,hiyama10,bodmer84}. 
However, the CM has not gone, for obvious computational reasons, beyond 
$_{\Lambda\Lambda}^{~11}{\rm Be}$. Furthermore, the CM has not been able to 
incorporate the full range of $\Lambda$ spin-dependent interactions that 
enter hypernuclear computations. The shell model provides a viable alternative 
\cite{mil10,gsd71,mgdd85}. In this work we demonstrated that the simple SM 
estimate Eq.~(\ref{eq:BLL}) for $\Lambda\Lambda$ separation energies 
$B_{\Lambda\Lambda}$ in $\Lambda\Lambda$ hypernuclei, in terms of $(2J+1)$ 
averaged g.s. doublet separation energies ${\overline B}_{\Lambda}$ 
in $\Lambda$ hypernuclei, works well for the known $\Lambda\Lambda$ 
hypernuclear species. The spectroscopic information required to devise 
the ${\overline B}_{\Lambda}$ input that dominates the estimate for 
$B_{\Lambda\Lambda}$ is now available through recent SM studies \cite{mil10} 
that derive the spin-dependent $\Lambda N$ interaction parameters 
from the observed $\gamma$ ray transitions in $p$-shell $\Lambda$ 
hypernuclei \cite{tamura10}. We estimate the precision of 
$B_{\Lambda\Lambda}$ values thus extracted to be about 0.2 MeV. 

It was acknowledged that the application of the SM to $\Lambda$ hypernuclei 
throughout the $p$ shell suffers from inability to reproduce correctly the 
g.s. $\Lambda$ separation energy in $_{\Lambda}^{9}{\rm Be}$ because of the 
loose structure of its particle unstable $^8$Be core. We have indicated a 
way to bypass this difficulty in $_{\Lambda\Lambda}^{~10}{\rm Be}$ by using 
an alternative SM estimate, Eq.~(\ref{eq:BLLalt}), that restores the desired 
level of predictability to the SM in this particular case. Using the uniquely 
assigned $_{\Lambda\Lambda}^{~~6}{\rm He}$ datum, we were able to derive good 
estimates for the $\Lambda\Lambda$ separation energies of the other two known 
species $_{\Lambda\Lambda}^{~10}{\rm Be}$ and $_{\Lambda\Lambda}^{~13}{\rm B}$ 
in terms of experimentally derived $\Lambda$ hypernuclear separation energies, 
augmented by SM $p$-shell systematics. Our predictions for 
$_{\Lambda\Lambda}^{~11}{\rm Be}$, $_{\Lambda\Lambda}^{~12}{\rm Be}$ and 
$_{\Lambda\Lambda}^{~12}{\rm B}$ suggest that whereas a 
$_{\Lambda\Lambda}^{~11}{\rm Be}$ or $_{\Lambda\Lambda}^{~12}{\rm B}$ 
interpretation for the KEK-E176 emulsion event \cite{aoki09} generally 
accepted as $_{\Lambda\Lambda}^{~13}{\rm B}$ cannot be excluded, the recently 
reported KEK-E373 HIDA event \cite{nakazawa10} is unlikely to fit a proper 
$_{\Lambda\Lambda}^{~11}{\rm Be}$ or $_{\Lambda\Lambda}^{~12}{\rm Be}$ 
assignment.

\section*{Acknowledgements} 

Useful discussions with Emiko Hiyama are gratefully acknowledged. A.G. thanks 
ECT$^{\ast}$ Director Achim Richter for hospitality in Fall 2010 when this 
work was conceived and acknowledges partial support by the EU initiative FP7, 
HadronPhysics2, under Project No. 227431. D.J.M. acknowledges the support by 
the U.S. DOE under Contract DE-AC02-98CH10886 with the Brookhaven National 
Laboratory.


\begin{thebibliography}{99} 

\bibitem{sbg93} J.~Schaffner, C.B.~Dover, A.~Gal, C.~Greiner, H.~St\"{o}cker, 
Phys. Rev. Lett. 71 (1993) 1328; see also J.~Schaffner-Bielich, A.~Gal, Phys. 
Rev. C 62 (2000) 034311. 

\bibitem{jsb08} J.~Schaffner-Bielich, Nucl. Phys. A 804 (2008) 309, 
Nucl. Phys. A 835 (2010) 279, and references therein.

\bibitem{hashim06} O.~Hashimoto, H.~Tamura, Prog. Part. Nucl. Phys. 57 (2006) 
564, and references therein.  

\bibitem{nakazawa10} K.~Nakazawa, H.~Takahashi, Prog. Theor. Phys. Suppl. 185 
(2010) 335, and references therein; K.~Nakazawa [KEK-E373], Nucl. Phys. A 835 
(2010) 207. 

\bibitem{takahashi01} H.~Takahashi, et al., Phys. Rev. Lett. 87 (2001) 212502. 

\bibitem{fg02a} I.N.~Filikhin, A.~Gal, Phys. Rev. C 65 (2002) 041001(R). 

\bibitem{fg02b} I.N.~Filikhin, A.~Gal, Nucl. Phys. A 707 (2002) 491; 
I.N.~Filikhin, A.~Gal, V.M.~Suslov, Phys. Rev. C 68 (2003) 024002.  

\bibitem{hiyama02} E.~Hiyama, M.~Kamimura, T.~Motoba, T.~Yamada, Y.~Yamamoto, 
Phys. Rev. C 66 (2002) 024007. 

\bibitem{shoeb04} M.~Shoeb, Phys. Rev. C 69 (2004) 054003. 

\bibitem{ubs04} Q.N.~Usmani, A.R.~Bodmer, B.~Sharma, Phys. Rev. C 70 (2004) 
061001(R). 

\bibitem{nemura05} H.~Nemura, S.~Shinmura, Y.~Akaishi, K.S.~Myint, Phys. Rev. 
Lett. 94 (2005) 202502. 

\bibitem{uh06} A.A.~Usmani, Z.~Hasan, Phys. Rev. C 74 (2006) 034320. 

\bibitem{hiyama10} E.~Hiyama, M.~Kamimura, Y.~Yamamoto, T.~Motoba, Phys. Rev. 
Lett. 104 (2010) 212502.  

\bibitem{tamura10} H.~Tamura, et al., Nucl. Phys. A 835 (2010) 3. 

\bibitem{mil10} D.J.~Millener, Nucl. Phys. A 835 (2010) 11; see also 
D.J.~Millener, arXiv:1011.0367 [nucl-th], and references therein. 

\bibitem{aoki09} S.~Aoki, et al. [KEK-E176], Nucl. Phys. A 828 (2009) 191, 
and references therein. 

\bibitem{gsd71} A.~Gal, J.M.~Soper, R.H.~Dalitz, Ann. Phys. 63 (1971) 53.

\bibitem{akaishi00} Y.~Akaishi, T.~Harada, S.~Shinmura, K.S.~Myint, Phys. Rev. 
Lett. 84 (2000) 3539. 

\bibitem{rijken99} Th.A.~Rijken, V.J.G.~Stoks, Y.~Yamamoto, Phys. Rev. C 59 
(1999) 21. 

\bibitem{davis05} D.H.~Davis, Nucl. Phys. A 754 (2005) 3c, and references 
therein. 

\bibitem{mgdd85} D.J.~Millener, A.~Gal, C.B.~Dover, R.H.~Dalitz, Phys. Rev. C 
31 (1985) 499. 

\bibitem{ahn01} J.K.~Ahn, et al. [KEK E373], AIP Conf. Proc. 594 (2001) 180; 
the listed $B_{\Lambda\Lambda}({_{\Lambda\Lambda}^{~10}{\rm Be}_{\rm g.s.}})$ 
value is obtained from the experimentally deduced value $11.90\pm 0.13$ MeV 
by adding 3.04 MeV for the $2^+$ excitation energy, assuming equal $2^+$ core 
excitation energies in $_{\Lambda}^{9}{\rm Be}$ and in 
$_{\Lambda\Lambda}^{~10}{\rm Be}$. 

\bibitem{danysz63} M.~Danysz, et al., Nucl. Phys. 49 (1963) 121; R.H.~Dalitz, 
et al., Proc. R. Soc. Lond. A 426 (1989) 1. 

\bibitem{aoki91} S.~Aoki, et al. [KEK E176], Prog. Theor. Phys. 85 (1991) 
1287; C.B.~Dover, D.J.~Millener, A.~Gal, D.H.~Davis, Phys. Rev. C 44 (1991) 
1905. 

\bibitem{ma10} Y.~Ma, et al., Nucl. Phys. A 835 (2010) 422. 

\bibitem{bodmer84} A.R.~Bodmer, Q.N.~Usmani, J.~Carlson, Nucl. Phys. A 422 
(1984) 510, in particular the update of Eq.~(24) which yields 
$B_{\Lambda\Lambda}({_{\Lambda\Lambda}^{~10}{\rm Be}})=14.35$ MeV. 


\end{thebibliography}
\end{document}